\documentclass[preprint,12pt]{elsarticle}

\usepackage{lineno,hyperref}

\begin{document}

\begin{frontmatter}

\title{Excited State Mass spectra and Regge trajectories of Bottom Baryons}



\author{Kaushal Thakkar$^{a,*}$\corref{mycorrespondingauthor}}
\cortext[Kaushal Thakkar]{Corresponding author}
\ead{kaushal2physics@gmail.com}
\author{Zalak Shah$^{b}$, Ajay Kumar Rai$^{b}$ and P. C. Vinodkumar$^c$}
\address{$^a$Department of Applied Sciences and Humanities, GIDC Degree Engineering college, Abrama, Navsari-396406, India\\
$^b$Department of Applied Physics, Sardar Vallabhbhai National Institute of Technology, Surat-395007, India\\
$^c$Department of physics, Sardar Patel University, V.V. Nagar, Anand-388120, India}

\begin{abstract}
We present the mass spectra of radial and orbital excited states of singly heavy bottom baryons; $\Sigma_{b}^{+}, \Sigma_{b}^-,  \Xi_b^-, \Xi_b^0, \Lambda_b^0$ and $\Omega_b^-$. The QCD motivated hypercentral quark model is employed for the three body description of baryons and the form  of confinement potential is hyper coulomb plus linear. The first order correction to the confinement potential is also incorporated in this work.  The semi-electronic decay of  $\Omega_b$ and $\Xi_b$ are calculated using the spectroscopic parameters of the baryons. The computed results are compared with other theoretical predictions as well as with the available experimental observations. The Regge trajectories are plotted in (n, $M^2$) plane.

\end{abstract}

\begin{keyword}
Baryons, Potential models, Hadron mass models and calculations
\end{keyword}

\end{frontmatter}
\section{Introduction}
The study of baryons containing heavy quarks (c, b) is rapidly growing due to the numerous number of data recently reported by various world wide experimental  facilities like LHCb, CDF, DO, CMS, SELEX etc., \cite{CDF2011,D02011,CDF2007,CDF22007,D02008,CDF2009,lhcb,101,cdf,103,104}. All ground states (with $J^{P}=\frac{1}{2}^{+}$, $\frac{3}{2}^{+}$) of singly charmed and bottom baryons (except $\Omega_{b}^*$) are reported experimentally \cite{olive}. At present, only two orbitally excited bottom baryons, $\Lambda_{b}^{0}$(5912) and $\Lambda_{b}^{0}$(5920) are experimentally known and reported  by LHCb Collaboration \cite{101}  with spin parity $J^{P}=\frac{1}{2}^{-}$ and  $J^{P}=\frac{3}{2}^{-}$ respectively. The excited states of other singly bottom baryons are expected to detect in near future.
\par Various phenomenological models have been used to study the baryons using different approaches. The theoretical predictions such as non-relativistic Isgur-Karl model \cite{Isgur}, relativized potential quark model \cite{Capstick}, relativistic quark model \cite{ebert2011}, the Fadeev approach \cite{13}, variational approach \cite{Roberts2008},  the chiral unitary model \cite{116}, non relativistic quantum mechanics \cite{woloshyn}, the extended local hidden gauge approach \cite{liang}, the relativistic flux tube (RFT) model \cite{chen2015}, the Hamiltonian model \cite{yoshida}, Regge phenomenology \cite
{kwei}, QCD sum rule \cite{ Hua, Yamaguchi2015, Mao, 117}, color hyperfine interaction \cite{118, Mkar}, Goldstone Boson Exchange Model \cite{ortega}, Soliton model \cite{90}  Quark-diquark model\cite{epjp, cjph} $\emph{etc}.,$ have been used to study the properties of heavy baryons. There are also many Lattice QCD studies which have examined the internal structure and quark dynamics of hadrons \cite{brown, mathur, 120}. However, there are limited efforts devoted to the study of radial and orbital excited states and decay properties of singly bottom baryons. Thus, presently, the theoretical studies of the excited states ($L\neq 0$) of singly bottom baryons  have become a subject of renewed interest. In our previous studies, we have calculated the radial and orbital excited state masses, semi-electronics decays, magnetic moments, etc. for singly charm baryons \cite{94,z}. In this paper, we extend the study for the mass spectra of singly bottom baryons and other decay properties.


\par The singly heavy baryons with one heavy quark (c or b) and two light quarks(u,d and s) give a perfect tool for studying the dynamics of the
light quarks in the presence of a heavy quark. The bottom baryons belong to two different SU(3) flavor representations:
$3 \otimes 3 = 6_{s} + \bar{3}_{A}$. The quark content and SU(3) multiplicity is mentioned in Table \ref{tab:1}. SU(3) symmetric sextet
and anti-symmetric anti-triplets are regulated as below \cite{crede}.
\begin{center}
$\Sigma_{b}^{+}$= $uub$,   $\Sigma_{b}^{0}$= $\frac{1}{\sqrt{2}}$ (ud +du)b,   $\Sigma_{b}^{-}$= ddb, $\Omega_{b}^{0}$= ssb\\
$\Lambda_{b}= \frac{1}{\sqrt{2}} (ud-du)b$,   $\Xi_{b}^{0}=\frac{1}{\sqrt{2}} (us-su)b$,
$\Xi_{b}^{-}=\frac{1}{\sqrt{2}} (ds-sd)b$
\end{center}
\begin{table*}
\begin{center}
\caption{\label{tab:1} List of heavy singly bottom baryons and their quark content.}
\begin{tabular}{lccc}
\hline\hline
Baryon & Quark Content& SU(3) multiplicity & (I, $I_{3}$) \\
\hline
$\Lambda_{b}^{0}$ & udb &  $\bar{3}$ & (0,0)\\
$\Sigma_{b}^{+}$ & uub & 6 &(1,1) \\
$\Sigma_{b}^{0}$ & udb & 6 &(1,0) \\
$\Sigma_{b}^{-}$ & ddb & 6 & (1,-1) \\
$\Xi_{b}^{0}$ & usb &$\bar{3}$ & ($\frac{1}{2}$,$\frac{1}{2}$)\\
$\Xi_{b}^{-}$ & dsb & $\bar{3}$ & ($\frac{1}{2}$,$-\frac{1}{2})$\\
$\Omega_{b}^{-}$ & ssb & 6 &(0,0) \\
\hline\hline
\end{tabular}

\end{center}
\end{table*}

\begin{table*}
\caption{\label{tab:2} The singly bottom baryon masses(MeV) with $J^{P}$ values as listed in PDG-2016 \cite{olive}.}
 \resizebox{\textwidth}{!}{
\begin{tabular}{cc|cc|cc|cc|c}
\hline\hline
Names & Mass & Names & Mass &  Names & Mass &  Names & Mass & $J^{P}$\\
\hline
$\Lambda_{b}(5619)^{0}$ &5619.5$\pm$.04 & $\Sigma_{b}(5811)^{+}$ &5811.3$\pm$1.7 & $\Xi_{b}(5790)^{-}$ & 5794.9 $\pm$ 0.9 & $\Omega_{b}(6048)^{-}$ & 6048.8 $\pm$ 3.2&$ \frac{1}{2}^{+}$\\

- &-& $\Sigma_{b}(5816)^{-}$ &5815.5$\pm$1.7 & $\Xi_{b}(5790)^{0}$ & 5793.1$\pm$2.5 &-&-&$ \frac{1}{2}^{+}$ \\
- & -&$\Sigma_{b}(5832)^{+}$ & 5832.1$\pm$0.7 & $\Xi_{b}(5945)^{0}$ &5949.3 $\pm$0.8 $\pm$ 0.9& - &- &$ \frac{3}{2}^{+}$\\
-&- & $\Sigma_{b}(5835)^{-}$ & 5835.1$\pm$0.6 &-  & - & - &- &$ \frac{3}{2}^{+}$\\
$\Lambda_{b}(5912)^{0}$ &5912.1$\pm$0.1$\pm$0.4&- &- &- &- &-&-&$\frac{1}{2}^{-}$\\
$\Lambda_{b}(5920)^{0}$ &5919.73$\pm$0.32 &- &- &- &- &-&-&$\frac{3}{2}^{-}$\\
\hline\hline
\end{tabular}}
\end{table*}
The experimentally known masses of singly bottom baryons are listed in Table \ref{tab:2}. The Hypercentral Constituent quark model is employed for the present study which has already been successfully used  for the study of baryons in light as well as heavy sector \cite{94,z,bhavin,15,93,95,EPJC}.

This paper is organized as follows: The hypercentral Constituent Quark Model (hCQM) applied for the study of singly bottom baryon mass spectroscopy presented in section 2. The mass spectra of bottom baryons are analysed and the Regge trajectories for the same are presented in section 3. The semi electronic weak decays of $\Xi_b$ and $\Omega_b$ baryons are computed and the details are presented in section 4.  In section 5, we have drawn important conclusions and summarized our present study on singly bottom baryons.

\section{Theoretical Framework: Hypercentral Constituent Quark Model (hCQM)}
The hypercentral approach has been applied to solve bound states and scattering problems in many different fields of physics and chemistry. The basic idea
 of the hypercentral approach to three-body systems is very simple. The two relative coordinates ($\vec{\rho}$ and $\vec{\lambda}$) are rewritten into a
 single six-dimensional vector and the nonrelativistic $Schr\ddot{o}dinger$ equation in the six-dimensional space is solved. The potential expressed in
 terms of the hypercentral radial co-ordinate, takes care of the three body interaction effectively. Such an attempt has already been employed for the mass
 spectra of singly heavy charmed baryons ($\Lambda_c^0$, $\Sigma_c^{++,+,0}$, $\Xi_c^{+,0}$ and $\Omega_c^0$)  and also for doubly heavy charmed baryons
 (both $\Omega$'s and $\Xi$'s families) in our previous work \cite{94, z, EPJC}. Details to this hypercentral constituent quark model employed for
 the present study of singly heavy bottom baryons is described below. The Jacobi coordinates to describe baryon as a bound state of three different
 constituent quarks are given by \cite{Bijker}.
\begin{equation}
\vec{\rho} = \frac{1}{\sqrt{2}}(\vec{r_{1}} - \vec{r_{2}})
\end{equation}
\begin{equation}
\vec{\lambda} =\frac{m_1\vec{r_1}+m_2\vec{r_2}-(m_1+m_2)\vec{r_3}}{\sqrt{m_1^2+m_2^2+(m_1+m_2)^2}}
\end{equation}
The respective reduced masses are given by
\begin{equation}
m_{\rho}=\frac{2 m_{1} m_{2}}{m_{1}+ m_{2}}
\end{equation}
\begin{equation}
 m_{\lambda}=\frac{2 m_{3} (m_{1}^2 + m_{2}^2+m_1m_2)}{(m_1+m_2)(m_{1}+ m_{2}+ m_{3})}
\end{equation}
Here, $m_1, m_2, m_3 $ are the constituent quark masses. We consider $m_{u}$= 0.338, $m_{d}$=0.350, $m_{s}$=0.500, $m_{b}$=4.67 (all in GeV). The angle of the Hyperspherical coordinates are given by $\Omega_{\rho}= (\theta_{\rho}, \phi_{\rho})$ and $\Omega_{\lambda}= (\theta_{\lambda}, \phi_{\lambda})$. We define hyper radius $x$ and hyper angle $\xi$ by,
\begin{equation}
x= \sqrt{\rho^{2} + \lambda^{2}}\,\,\,and\,\,\, \xi= arctan \left(\frac{\rho}{\lambda} \right)
\end{equation}
\begin{table}[]
\begin{center}
\caption{\label{tab:table4}Quark mass parameters (in GeV) and constants used in the calculations.}
\begin{tabular}{cccccccc}
\hline\hline
${m_{u}}$ &${m_{d}}$& ${m_{s}}$& ${m_{b}}$ & $C_{F}$ & $C_{A}$ &$n_{f}$ & $\alpha_s(\mu_{0}$=1 GeV)\\
\hline
0.338 & 0.350 &0.500 &4.67 & $\frac{2}{3}$ &3 &5 &0.6\\
\hline
\end{tabular}
\end{center}
\end{table}
In the center of mass frame ($R_{c.m.} = 0$), the kinetic energy operator can be written as
\begin{equation}
\frac{P_{x}^2}{2m}=-\frac{\hbar^2}{2m}(\bigtriangleup_{\rho} + \bigtriangleup_{\lambda})= -\frac{\hbar^2}{2m}\left(\frac{\partial^2}{\partial x^2}+\frac{5}{x}\frac{\partial}{\partial x}+\frac{L^2(\Omega)}{x^2}\right)
\end{equation}
where $m$=$\frac{2 m_{\rho}m_{\lambda}}{m_{\rho}+m_{\lambda}}$ is the reduced mass and $L^2(\Omega)$=$L^2(\Omega_{\rho},\Omega_{\lambda},\xi)$ is the quadratic Casimir operator of the six-dimensional rotational group O(6) and its eigenfunctions are the hyperspherical harmonics, $Y_{[\gamma]l_{\rho}l_{\lambda}}$ ($\Omega_{\rho}$,$\Omega_{\lambda}$,$\xi$) satisfying the eigenvalue relation,
 $L^2Y_{[\gamma]l_{\rho}l_{\lambda}}$
($\Omega_{\rho}$,$\Omega_{\lambda},\xi)$=-$\gamma (\gamma +4) Y_{[\gamma]l_{\rho}l_{\lambda}}(\Omega_{\rho},\Omega_{\lambda},\xi)$. Here, $\vec{L}=\vec{L_{\rho}}+\vec{L_{\lambda}}$, $l_{\rho}$ and $l_{\lambda}$ are the angular momenta associated with the $\vec{\rho}$ and $\vec{\lambda}$
variables respectively and $\gamma$ is the hyper angular momentum quantum number.
\par The confining three-body potential is chosen within a string-like picture, where the quarks are connected by gluonic strings and the potential increases linearly with a collective radius $x$ as mentioned in \cite{ginnani2015,M. Ferraris}. In the hypercentral approximation, the potential is expressed in terms of the hyper radius ($x$) as
\begin{equation}
\sum_{i<j}V(r_{ij})=V(x)+. . . .
\end{equation}
In this case the potential $V(x)$ not only contains two-body interactions but it contains three-body
effects also. The three-body effects are desirable in the study of hadrons since the non-
Abelian nature of QCD leads to gluon-gluon couplings which produce three-body forces.

The  model Hamiltonian for baryons in the hCQM is then expressed as
\begin{equation}
H= \frac{P_{x}^{2}}{2m} +V(x)
\end{equation}
The exact solution of the QCD equations is very complex, so one has to rely upon conventional quark models. The assumptions in various conventional quark models are different, but they have a simple general structure in common including some basic features like confinement and asymptotic freedom and for the rest built up by means of suitable assumptions. More details on similarities and differences in various quark models can be found in Ref. \cite{ginnani2015,Giannini1990}. The main differences between Hamiltonian used in Hypercentral Constituent Quark Model (HCQM) adopted in this paper and conventional quark model given by Isgur and Karl \cite{Isgur} are as follow.
\begin{enumerate}
\item	{The confinement potential used in Isgur and Karl quark model is given by harmonic oscillator plus constant potential, while the confinement potential used in HCQM is given by linear plus hyper coulomb potential.}

\item	{The mass of the light quarks (u and d) were same (m1=m2$\neq$m3) in Isgur and Karl quark model, while in this paper, we have used unequal quark masses (m1$\neq$m2$\neq$m3) in HCQM.}

\item	{In Isgur and Karl quark model only hyperfine part is kept as a spin dependent potential, while here we have used spin-spin, spin-orbit as well as tensor terms as a spin dependent potential.}

\item	{We have solved Schrodinger equation in six dimensional space for HCQM, while in conventional quark model like Isgur and Karl, Schrodinger equation is solved in three dimensional space.}

\end{enumerate}

\begin{table}
\begin{center}

\caption{\label{tab:4}Mass spectra of $\Lambda_{b}^{0}$ baryon (in GeV).}

\resizebox{\textwidth}{!}{
\begin{tabular}{cc|cccccccccc}
\hline\hline
State&$J^P$ & A & B & Exp. \cite{olive} & \cite{ebert2011} & \cite{yoshida}& \cite{chen2015}& \cite{118} &\cite{kwei}& \cite{liang}&\cite{Yamaguchi2015}\\
n$^{2S+1} L_{J}$\\
\hline
$(1^2S_{1/2})$&$\frac{1}{2}^+$&5.621&5.621&5.620&5.620&5.618&5.619&&5.619&&5.612\\
$(2^2S_{1/2})$&$\frac{1}{2}^+$&6.016&6.026&&6.089&&&&&&6.107\\
$(3^2S_{1/2})$&$\frac{1}{2}^+$&6.364&6.380&&6.455&&&&&&6.338\\
$(4^2S_{1/2})$&$\frac{1}{2}^+$&6.697&6.719&&6.756\\
$(5^2S_{1/2})$&$\frac{1}{2}^+$&7.022&7.050&&7.015\\
$(6^2S_{1/2})$&$\frac{1}{2}^+$&7.343&7.377&&7.256\\
\hline
$(1^2P_{1/2})$&$\frac{1}{2}^-$&	5.992	&	6.000	&	5.912 	&	5.930 &5.938& 5.911&5.929&&5.820	\\
$(1^2P_{3/2})$&$\frac{3}{2}^-$	&	5.980	&	5.988	&5.920 	&5.942&5.939	&5.920&5.940&5.913&5.969	\\
\hline
$(2^2P_{1/2})$&$\frac{1}{2}^-$&	6.303	&	6.317	&		&	6.326&6.236	\\
$(2^2P_{3/2})$&$\frac{3}{2}^-$&	6.290	&	6.304	&		&	6.333&6.273	\\
	\hline	
$(3^2P_{1/2})$&$\frac{1}{2}^-$&	6.615	&	6.635	&		&	6.645&6.273	\\
$(3^2P_{3/2})$&$\frac{3}{2}^-$&	6.602	&	6.621	&		&	6.651&6.285	\\
		\hline	
$(4^2P_{1/2})$&$\frac{1}{2}^-$&		6.928	&	6.954	&		&	6.917	\\
$(4^2P_{3/2})$&$\frac{3}{2}^-$&		6.914	&	6.939	&		&	6.922	\\
\hline
$(5^2P_{1/2})$&$\frac{1}{2}^-$&	7.241	&	7.273	&		&		7.157\\
$(5^2P_{3/2})$&$\frac{3}{2}^-$&	7.226	&	7.258	&		&	7.171	\\
		\hline	
$(1^2D_{3/2})$&$\frac{3}{2}^+$	&	6.219	&	6.233	&		&	6.190& &6.147	\\
$(1^2D_{5/2})$&$\frac{5}{2}^+$	&	6.200	&	6.213	&		&6.196	&6.212	&6.153&&6.193\\
	\hline	
$(2^2D_{3/2})$&$\frac{3}{2}^+$	&	6.528	&	6.548	&		&	6.526	\\
$(2^2D_{5/2})$&$\frac{5}{2}^+$	&	6.509	&	6.527	&		&	6.531&6.530	\\
	\hline	
$(3^2D_{3/2})$&$\frac{3}{2}^+$	&	6.839	&	6.865	&		&	6.811	\\
$(3^2D_{5/2})$&$\frac{5}{2}^+$	&	6.819	&	6.844	&		&	6.814	\\
	\hline	
$(4^2D_{3/2})$&$\frac{3}{2}^+$	&	7.151	&	7.183	&		&	7.060	\\
$(4^2D_{5/2})$&$\frac{5}{2}^+$	&	7.129	&	7.160	&		&	7.063	\\
	\hline	
$(1^2F_{5/2})$&$\frac{5}{2}^-$	&	6.442	&	6.460	&		&	6.408&&6.346	\\
$(1^2F_{7/2})$&$\frac{7}{2}^-$	&	6.415	&	6.432	&		&	6.411&&6.351&&6.461	\\
	\hline	
$(2^2F_{5/2})$&$\frac{5}{2}^-$	&	6.750	&	6.775	&		&	6.705	\\
$(2^2F_{7/2})$&$\frac{7}{2}^-$	&	6.723	&	6.747	&		&	6.708	\\
	\hline	
$(3^2F_{5/2})$&$\frac{5}{2}^-$	&	7.058	&	7.090	&		&	6.964	\\
$(3^2F_{7/2})$&$\frac{7}{2}^-$	&	7.033	&	7.062	&		&	6.966	\\
		\hline	
$(4^2F_{5/2})$&$\frac{5}{2}^-$	&	7.371	&	7.408	&		&	7.196	\\
$(4^2F_{7/2})$&$\frac{7}{2}^-$	&	7.343	&	7.379	&		&	7.197	\\
\hline\hline
\end{tabular}
}
\end{center}
\end{table}

\begin{table}
\begin{center}
\caption{\label{tab:7} Mass spectra of $\Sigma_{b}$ baryons (in GeV).}
\scalebox{0.6}{
\begin{tabular}{cc|cccccccccccc}
\hline\hline
&&\multicolumn{2}{c}{$\Sigma_{b}^{-}$} &\multicolumn{2}{c}{$\Sigma_{b}^{+}$}&\multicolumn{5}{c}{Refs.}&&\\
\cline{3-4}
\cline{5-6}
\cline{7-12}
State&$J^P$& A & B & A & B&\cite{ebert2011}&\cite{yoshida}&\cite{kwei}&\cite{Mkar}& \cite{Yamaguchi2015}&Others \\
\hline

$(1^2S_{1/2})$&$\frac{1}{2}^+$&5.816&5.816&5.811&5.811&5.808&5.823&5.813&&5.833&5.814\cite{118}\\
$(2^2S_{1/2})$&$\frac{1}{2}^+$&6.256&6.262&6.269&6.275&6.213&&&&6.294\\\
$(3^2S_{1/2})$&$\frac{1}{2}^+$&6.627&6.641&6.665&6.669&6.575&&&&6.447\\
$(4^2S_{1/2})$&$\frac{1}{2}^+$&6.997&7.018&7.041&7.061&6.869\\
$(5^2S_{1/2})$&$\frac{1}{2}^+$&7.367&7.394&7.425&7.452&7.124\\
\hline
$(1^4S_{3/2})$&$\frac{3}{2}^+$&5.835&5.835&5.832&5.832&5.834&5.845&5.833&&5.858&5.858\cite{118}\\
$(2^4S_{3/2})$&$\frac{3}{2}^+$&6.271&6.277&6.285&6.291&6.226&&&&6.326\\
$(3^4S_{3/2})$&$\frac{3}{2}^+$&6.637&6.650&6.666&6.679&6.583&&&&6.447\\
$(4^4S_{3/2})$&$\frac{3}{2}^+$&7.004&7.024&7.048&7.067&6.876\\
$(5^4S_{3/2})$&$\frac{3}{2}^+$&7.371&7.398&7.430&7.457&7.129\\
\hline
$(1^2P_{1/2})$&$\frac{1}{2}^-$&	6.122	&	6.129&6.131	&	6.139&6.101&	6.127&&6.095&&6.00 \cite{87}	\\
$(1^2P_{3/2})$&$\frac{3}{2}^-$&	6.105	&	6.112&6.113	&	6.120&	6.096&6.132&6.098&6.101&& 5.91\cite{92}	\\
$(1^4P_{1/2})$&$\frac{1}{2}^-$&	6.130&	6.137	&6.140	&	6.148	&6.095&	&&6.087	\\
$(1^4P_{3/2})$&$\frac{3}{2}^-$&	6.113	&	6.121&	6.122	&	6.129&6.087&&&6.096	\\
$(1^4P_{5/2})$&$\frac{5}{2}^-$&	6.092	&	6.099	&	6.098	&	6.104&6.084	&6.144&6.117&6.084	\\
\hline
$(2^2P_{1/2})$&$\frac{1}{2}^-$&	6.487	&	6.502&	6.512	&	6.525	&6.440&6.135\\
$(2^2P_{3/2})$&$\frac{3}{2}^-$&	6.470	&	6.484	&	6.492	&	6.506&6.430&6.141	\\
$(2^4P_{1/2})$&$\frac{1}{2}^-$&	6.496	&	6.512&	6.522	&	6.535	&6.430\\
$(2^4P_{3/2})$&$\frac{3}{2}^-$&	6.479	&	6.493	&	6.502	&	6.515&6.423	\\
$(2^4P_{5/2})$&$\frac{5}{2}^-$&	6.455	&	6.468&	6.476	&	6.489&6.421&6.592	\\
\hline
$(3^2P_{1/2})$&$\frac{1}{2}^-$&	6.855	&	6.875	&	6.894	&	6.915&6.756	&6.246	\\
$(3^2P_{3/2})$&$\frac{3}{2}^-$&	6.836	&	6.856	&	6.873	&	6.893&6.742	&6.246	\\
$(3^4P_{1/2})$&$\frac{1}{2}^-$&	6.865	&	6.884	&	6.904	&	6.926&6.096	&&&	\\
$(3^4P_{3/2})$&$\frac{3}{2}^-$&	6.845	&	6.866	&	6.884	&	6.904&6.736		&&&	\\
$(3^4P_{5/2})$&$\frac{5}{2}^-$&	6.820	&	6.841	&	6.856	&	6.875&6.732	&6.834&	&	\\
\hline
$(4^2P_{1/2})$&$\frac{1}{2}^-$&	7.221	&	7.250	&	7.277	&	7.305&7.024&&&	\\
$(4^2P_{3/2})$&$\frac{3}{2}^-$&	7.202	&	7.230	&	7.255	&	7.282&7.009&&	&	\\
$(4^4P_{1/2})$&$\frac{1}{2}^-$&	7.230	&	7.261	&	7.288	&	7.316&7.008&&	&	\\
$(4^4P_{3/2})$&$\frac{3}{2}^-$&	7.212	&	7.240	&	7.266	&	7.293&	7.003&&	&	\\
$(4^4P_{5/2})$&$\frac{5}{2}^-$&	7.187	&	7.213	&	7.236	&	7.262&6.999&&&	\\
\hline
$(5^2P_{1/2})$&$\frac{1}{2}^-$&	7.590	&	7.625			&	7.660	&	7.694	\\
$(5^2P_{3/2})$&$\frac{3}{2}^-$&	7.570	&	7.604			&	7.637	&	7.670	\\
$(5^4P_{1/2})$&$\frac{1}{2}^-$&	7.601	&	7.636			&	7.671	&	7.705	\\
$(5^4P_{3/2})$&$\frac{3}{2}^-$&	7.580	&	7.614		&	7.648	&	7.682	\\
$(5^4P_{5/2})$&$\frac{5}{2}^-$&	7.552	&	7.586		&	7.618	&	7.651	\\
\hline
$(1^2D_{3/2})$&$\frac{3}{2}^+$&	6.389	&	6.403	&	6.409	&	6.423&6.326		&&&	\\
$(1^4D_{3/2})$&$\frac{3}{2}^+$&	6.399	&	6.414	&	6.420	&	6.434&6.285		&&&	\\
$(1^4D_{1/2})$&$\frac{1}{2}^+$&	6.418	&	6.434	&6.441	&	6.457&6.311	&&	&	\\
$(1^4D_{5/2})$&$\frac{5}{2}^+$&	6.373	&	6.386	&	6.391	&	6.405&6.270		&&&	\\
$(1^2D_{5/2})$&$\frac{5}{2}^+$&	6.363	&	6.376	&6.381&	6.394&6.284&6.397		&6.369	\\
$(1^4D_{7/2})$&$\frac{7}{2}^+$&	6.340	&	6.352	&	6.441	&	6.368&6.260	&&6.388	\\
\hline
$(2^2D_{3/2})$&$\frac{3}{2}^+$&	6.753	&	6.774	&6.787	&	6.809&6.647		&&&		\\
$(2^4D_{3/2})$&$\frac{3}{2}^+$&	6.763	&	6.784	&	6.797	&	6.821&6.612		&&&	\\
$(2^4D_{1/2})$&$\frac{1}{2}^+$&	6.784	&	6.805	&6.819	&	6.844&6.636	&&	&	\\
$(2^4D_{5/2})$&$\frac{5}{2}^+$&	6.736	&	6.756	&	6.769	&	6.790&6.598	&&&	\\
$(2^2D_{5/2})$&$\frac{5}{2}^+$&	6.726	&	6.746	&	6.758	&	6.778&6.612	&6.402&&	\\
$(2^4D_{7/2})$&$\frac{7}{2}^+$&	6.702	&	6.721&6.733	&	6.751	&6.590		&&&		\\
\hline
$(1^4F_{3/2})$&$\frac{3}{2}^-$&	6.690	&	6.714	&	6.725	&	6.749&6.550	&&&	\\
$(1^2F_{5/2})$&$\frac{5}{2}^-$&	6.651	&	6.672	&	6.682	&	6.703&6.564	&&&	\\
$(1^4F_{5/2})$&$\frac{5}{2}^-$&	6.661	&	6.683	&	6.694	&	6.715&6.501	&&&	\\
$(1^4F_{7/2})$&$\frac{7}{2}^-$&	6.626	&	6.645	&	6.654	&	6.673&6.472	&&&	\\
$(1^2F_{7/2})$&$\frac{7}{2}^-$&	6.615	&	6.633	&	6.643	&	6.661	&6.500&&6.630	&&&\\
$(1^4F_{9/2})$&$\frac{9}{2}^-$&	6.583	&	6.599	&	6.607	&	6.610&6.459&&6.648	\\
\hline\hline
\end{tabular}
}

\end{center}
\end{table}

\begin{table}
\begin{center}
\caption{\label{tab:11}Mass spectra of $\Xi_{b}$ baryons (in GeV).}
\scalebox{0.6}{
\begin{tabular}{cc|ccccccccc}
\hline\hline
&&\multicolumn{2}{c}{$\Xi_{b}^{0}$} &\multicolumn{2}{c}{$\Xi_{b}^{-}$}&\multicolumn{5}{c}{Refs.}\\
\cline{2-3}
\cline{4-5}
\cline{6-10}
State&$J^P$& A & B  & A & B &\cite{ebert2011} & \cite{118}& \cite{Roberts2008}&\cite{kwei}& Others\\
\hline
$(1^2S_{1/2})$&$\frac{1}{2}^+$&5.793&5.793&5.795&5.795&5.803&5.795&5.806&5.793&5.806\cite{Yamaguchi2015}\\
$(2^2S_{1/2})$&$\frac{1}{2}^+$&6.193&6.203&6.180&6.189&6.266\\
$(3^2S_{1/2})$&$\frac{1}{2}^+$&6.537&6.554&6.512&6.527&6.601\\
$(4^2S_{1/2})$&$\frac{1}{2}^+$&6.868&6.892&6.831&6.853&6.913\\
$(5^2S_{1/2})$&$\frac{1}{2}^+$&7.93&7.223&7.145&7.172&7.165\\
\hline
$(1^4S_{3/2})$&$\frac{3}{2}^+$&5.948&5.948&5.942&5.942&&&5.980&5.952&5.929\cite{117}\\
$(2^4S_{3/2})$&$\frac{3}{2}^+$&6.309&6.316&6.292&6.298\\
$(3^4S_{3/2})$&$\frac{3}{2}^+$&6.611&6.625&6.583&6.595\\
$(4^4S_{3/2})$&$\frac{3}{2}^+$&6.919&6.940&6.880&6.899\\
$(5^4S_{3/2})$&$\frac{3}{2}^+$&7.231&7.258&7.181&7.206\\
\hline
$(1^2P_{1/2})$&$\frac{1}{2}^-$&	6.143	&	6.151		&	6.131	&	6.139&	6.120& 6.106&6.090&&6.097\cite{chen2015}	\\
$(1^2P_{3/2})$&$\frac{3}{2}^-$&	6.133	&	6.141		&	6.122	&	6.129&	6.130&6.115&6.093&6.080&6.106 \cite{chen2015}	\\
$(1^4P_{1/2})$&$\frac{1}{2}^-$&	6.133	&	6.141			&	6.122	&	6.129& &&&&6.140 \cite{87}	\\
$(1^4P_{3/2})$&$\frac{3}{2}^-$&	6.138	&	6.146			&	6.126	&	6.134&&&&&6.06 \cite{92}	\\
$(1^4P_{5/2})$&$\frac{5}{2}^-$&	6.124	&	6.132			&	6.114	&	6.121&&&&6.232	\\
\hline
$(2^2P_{1/2})$&$\frac{1}{2}^-$&	6.457	&	6.472		&	6.434	&	6.448&	6.496	\\
$(2^2P_{3/2})$&$\frac{3}{2}^-$&	6.446	&	6.460		&	6.423	&	6.437&	6.502	\\
$(2^4P_{1/2})$&$\frac{1}{2}^-$&	6.462	&	6.478			&	6.439	&	6.453	\\
$(2^4P_{3/2})$&$\frac{3}{2}^-$&	6.451	&	6.466			&	6.429	&	6.442	\\
$(2^4P_{5/2})$&$\frac{5}{2}^-$&	6.436	&	6.451			&	6.415	&	6.428	\\
\hline
$(3^2P_{1/2})$&$\frac{1}{2}^-$&	6.770	&	6.792		&	6.737	&	6.757&	6.805	\\
$(3^2P_{3/2})$&$\frac{3}{2}^-$&	6.759	&	6.781		&	6.726	&	6.746&	6.810	\\
$(3^4P_{1/2})$&$\frac{1}{2}^-$&	6.776	&	6.798			&	6.743	&	6.762	\\
$(3^4P_{3/2})$&$\frac{3}{2}^-$&	6.765	&	6.786			&	6.732	&	6.751	\\
$(3^4P_{5/2})$&$\frac{5}{2}^-$&	6.750	&	6.771			&	6.717	&	6.737	\\
\hline
$(4^2P_{1/2})$&$\frac{1}{2}^-$&	7.087	&	7.115		&	7.041	&	7.068&	7.068	\\
$(4^2P_{3/2})$&$\frac{3}{2}^-$&	7.074	&	7.102		&	7.030	&	7.056&	7.073	\\
$(4^4P_{1/2})$&$\frac{1}{2}^-$&	7.093	&	7.122			&	7.047	&	7.074	\\
$(4^4P_{3/2})$&$\frac{3}{2}^-$&	7.080	&	7.109			&	7.036	&	7.062	\\
$(4^4P_{5/2})$&$\frac{5}{2}^-$&	7.064	&	7.091			&	7.020	&	7.046	\\
\hline
$(5^2P_{1/2})$&$\frac{1}{2}^-$&7.402	&	7.437		&	7.345	&	7.378&	7.302	\\
$(5^2P_{3/2})$&$\frac{3}{2}^-$&7.389	&	7.424		&	7.333	&	7.366&	7.306	\\
$(5^4P_{1/2})$&$\frac{1}{2}^-$&7.408	&	7.443		&	7.351	&	7.384	\\
$(5^4P_{3/2})$&$\frac{3}{2}^-$&7.395	&	7.430		&	7.339	&	7.372	\\
$(5^4P_{5/2})$&$\frac{5}{2}^-$&7.378	&	7.413		&	7.324	&	7.356	\\
\hline
$(1^2D_{3/2})$&$\frac{3}{2}^+$&	6.371	&	6.386		&	6.350	&	6.364&	6.366&6.344	&&&6.190\cite{Hua}\\
$(1^4D_{3/2})$&$\frac{3}{2}^+$&	6.377	&	6.392		&	6.356	&	6.370	\\
$(1^4D_{1/2})$&$\frac{1}{2}^+$&	6.389	&	6.405		&	6.366	&	6.382	\\
$(1^4D_{5/2})$&$\frac{5}{2}^+$&	6.361	&	6.375	&	6.341	&	6.355	\\
$(1^2D_{5/2})$&$\frac{5}{2}^+$&	6.355	&	6.369	&	6.336	&	6.349&	6.373&6.349&&6.354	\\
$(1^4D_{7/2})$&$\frac{7}{2}^+$&	6.341	&	6.354		&	6.323	&	6.335&&&&6.499	\\
\hline
$(2^2D_{3/2})$&$\frac{3}{2}^+$&	6.682	&	6.704		&	6.652	&	6.671&	6.690	\\
$(2^4D_{3/2})$&$\frac{3}{2}^+$&	6.689	&	6.710			&	6.657	&	6.677	\\
$(2^4D_{1/2})$&$\frac{1}{2}^+$&	6.701	&	6.723			&	6.669	&	6.689	\\
$(2^4D_{5/2})$&$\frac{5}{2}^+$&	6.672	&	6.693			&	6.642	&	6.661	\\
$(2^2D_{5/2})$&$\frac{5}{2}^+$&	6.666	&	6.687		&	6.636	&	6.655&	6.696	\\
$(2^4D_{7/2})$&$\frac{7}{2}^+$&	6.652	&	6.672		&	6.623	&	6.661	\\
\hline
$(1^4F_{3/2})$&$\frac{3}{2}^-$&	6.620	&	6.642		&	6.589	&	6.610	\\
$(1^2F_{5/2})$&$\frac{5}{2}^-$&	6.595	&	6.615		&	6.566	&	6.585&	6.577&6.555	\\
$(1^4F_{5/2})$&$\frac{5}{2}^-$&	6.602	&	6.622		&	6.573	&	6.592	\\
$(1^4F_{7/2})$&$\frac{7}{2}^-$&	6.579	&	6.598			&	6.552	&	6.570	\\
$(1^2F_{7/2})$&$\frac{7}{2}^-$&	6.572	&	6.591	&	6.545	&	6.563&	6.581&6.559&&6.616	\\
$(1^4F_{9/2})$&$\frac{9}{2}^-$&	6.551	&	6.569			&	6.527	&	6.543&&&&6.756	\\
\hline\hline
\end{tabular}
}
\end{center}
\end{table}

\begin{table}
\begin{center}
\caption{\label{tab:9} Mass spectra of $\Omega_{b}$ baryons (in GeV).}
\scalebox{0.6}{
\begin{tabular}{cc|ccccccc}
\hline\hline
State&$J^P$& A & B & \cite{ebert2011} & \cite{yoshida}&\cite{kwei}& \cite{Yamaguchi2015}&\cite{Agaev2017} \\
\hline
$(1^2S_{1/2})$&$\frac{1}{2}^+$&6.048	&	6.048&6.054&6.076&6.048&6.081&6.024\\
$(2^2S_{1/2})$&$\frac{1}{2}^+$&6.448&6.455&6.450&&&6.472&6.325\\
$(3^2S_{1/2})$&$\frac{1}{2}^+$&6.786&6.799&6.804&&&6.593\\
$(4^2S_{1/2})$&$\frac{1}{2}^+$&7.120&7.140&7.091\\
$(5^2S_{1/2})$&$\frac{1}{2}^+$&7.453&7.480&7.338\\
\hline
$(1^4S_{3/2})$&$\frac{3}{2}^+$&6.086	&	6.086&6.088&6.094&&6.102&6.084\\
$(2^4S_{3/2})$&$\frac{3}{2}^+$&6.474&6.481&6.461&&&6.478&6.412\\
$(3^4S_{3/2})$&$\frac{3}{2}^+$&6.802&6.815&6.811&&&6.593\\
$(4^4S_{3/2})$&$\frac{3}{2}^+$&7.131&7.150&7.096\\
$(5^4S_{3/2})$&$\frac{3}{2}^+$&7.461&7.487&7.343\\
\hline
$(1^2P_{1/2})$&$\frac{1}{2}^-$& 6.331	&	6.338	& 6.339& 6.333& \\
$(1^2P_{3/2})$&$\frac{3}{2}^-$&	6.321	&	6.328	&6.340& 6.336&6.325\\
$(1^4P_{1/2})$&$\frac{1}{2}^-$&	6.336	&	6.343	&6.330&\\
$(1^4P_{3/2})$&$\frac{3}{2}^-$&	6.326	&	6.333	&6.331&\\
$(1^4P_{5/2})$&$\frac{5}{2}^-$&	6.313	&	6.320	&6.334&6.345\\
\hline
$(2^2P_{1/2})$&$\frac{1}{2}^-$&	6.659	&	6.673	&6.710&6.340	\\
$(2^2P_{3/2})$&$\frac{3}{2}^-$&	6.649	&	6.662	&6.705&6.344	\\
$(2^4P_{1/2})$&$\frac{1}{2}^-$&	6.665	&	6.679	&6.706	\\
$(2^4P_{3/2})$&$\frac{3}{2}^-$&	6.654	&	6.668	&6.699	\\
$(2^4P_{5/2})$&$\frac{5}{2}^-$&	6.640	&	6.653	&6.700&6.728	\\
\hline
$(3^2P_{1/2})$&$\frac{1}{2}^-$&	6.988	&	7.009	&7.009	&6.437\\
$(3^2P_{3/2})$&$\frac{3}{2}^-$&	6.977	&	6.998	&7.002&6.437	\\
$(3^4P_{1/2})$&$\frac{1}{2}^-$&	6.993	&	7.015	&7.003	\\
$(3^4P_{3/2})$&$\frac{3}{2}^-$&	6.983	&	7.003	&6.998	\\
$(3^4P_{5/2})$&$\frac{5}{2}^-$&	6.969	&	6.988	&6.996&6.919	\\
\hline
$(4^2P_{1/2})$&$\frac{1}{2}^-$&	7.319	&	7.346	&7.265	\\
$(4^2P_{3/2})$&$\frac{3}{2}^-$&	7.307	&	7.334	&7.258	\\
$(4^4P_{1/2})$&$\frac{1}{2}^-$&	7.325	&	7.352	&7.257	\\
$(4^4P_{3/2})$&$\frac{3}{2}^-$&	7.313	&	7.340	&7.250	\\
$(4^4P_{5/2})$&$\frac{5}{2}^-$&	7.297	&	7.323	&7.251	\\
\hline
$(5^2P_{1/2})$&$\frac{1}{2}^-$&	7.649	&	7.682	&	\\
$(5^2P_{3/2})$&$\frac{3}{2}^-$&	7.637	&	7.670	&	\\
$(5^4P_{1/2})$&$\frac{1}{2}^-$&	7.654	&	7.688	&	\\
$(5^4P_{3/2})$&$\frac{3}{2}^-$&	7.643	&	7.676	&	\\
$(5^4P_{5/2})$&$\frac{5}{2}^-$&	7.627	&	7.659	&	\\
\hline
$(1^2D_{3/2})$&$\frac{3}{2}^+$&	6.569	&	6.583	&6.549	\\
$(1^4D_{3/2})$&$\frac{3}{2}^+$&	6.574	&	6.589	&6.530	\\
$(1^4D_{1/2})$&$\frac{1}{2}^+$&	6.585	&	6.601	&6.540	\\
$(1^4D_{5/2})$&$\frac{5}{2}^+$&	6.560	&	6.573	&6.520	\\
$(1^2D_{5/2})$&$\frac{5}{2}^+$&	6.554	&	6.567	&6.529&6.561&6.590	\\
$(1^4D_{7/2})$&$\frac{7}{2}^+$&	6.541	&	6.553	&6.517&&6.609	\\
\hline
$(2^2D_{3/2})$&$\frac{3}{2}^+$&	6.896	&	6.678	&6.863	\\
$(2^4D_{3/2})$&$\frac{3}{2}^+$&	6.902	&	6.685	&6.846	\\
$(2^4D_{1/2})$&$\frac{1}{2}^+$&	6.914	&	6.699	&6.857	\\
$(2^4D_{5/2})$&$\frac{5}{2}^+$&	6.886	&	6.666	&6.837	\\
$(2^2D_{5/2})$&$\frac{5}{2}^+$&	6.880	&	6.659	&6.846&6.566	\\
$(2^4D_{7/2})$&$\frac{7}{2}^+$&	6.866	&	6.643	&6.834	\\
\hline
$(1^4F_{3/2})$&$\frac{3}{2}^-$&	6.826	&	6.846	&6.763	\\
$(1^2F_{5/2})$&$\frac{5}{2}^-$&	6.803	&	6.822	&6.771	\\
$(1^4F_{5/2})$&$\frac{5}{2}^-$&	6.809	&	6.828	&6.737	\\
$(1^4F_{7/2})$&$\frac{7}{2}^-$&	6.788	&	6.806	&6.736	\\
$(1^2F_{7/2})$&$\frac{7}{2}^-$&	6.782	&	6.800	&6.719&&6.844	\\
$(1^4F_{9/2})$&$\frac{9}{2}^-$&	6.763	&	6.780	&6.713&&6.863	\\
\hline\hline
\end{tabular}}

\end{center}
\end{table}

The six-dimensional hyperradial $Schr\ddot{o}dinger$ equation corresponds to the above Hamiltonian reduces to
\begin{equation}\label{eq:6}
\left[\frac{-1}{2m}\frac{d^{2}}{d x^{2}} + \frac{\frac{15}{4}+ \gamma(\gamma+4)}{2mx^{2}}+ V(x)\right]\phi_{ \gamma}(x)= E\phi_{\gamma}(x)
\end{equation}
where $\phi_{ \gamma}(x)$ is the reduced hypercentral wave function. If we compare above equation with the usual three-dimensional
radial $Schr\ddot{o}dinger$ equation, the resemblance between the angular momentum and the hyperangular momentum
is given by $l(l+1) \rightarrow \frac{15}{4} + \gamma(\gamma +4)$.

For the present study, we consider the hypercentral potential $V(x)$ as the hyper Coulomb plus linear potential with first order correction \cite{koma,11,20} and spin-dependent interaction, which is given as
\begin{equation}\label{eq:7}
V(x) =  V^{0}(x) + \left(\frac{1}{m_{\rho}}+ \frac{1}{m_{\lambda}}\right) V^{(1)}(x)+V_{SD}(x)
\end{equation}
where, $V^{0}(x)$ is defined as
\begin{equation}
V^{(0)}(x)= \frac{\tau}{x}+ \beta x
\end{equation}

Here, the hyper-Coulomb strength $\tau=-\frac{2}{3} \alpha_s$, $\frac{2}{3}$ is the color factor for the baryon. $\beta$ corresponds to the string tension of the confinement. We fix the model parameter $\beta$ to get the experimental spin average mass of the each ground state bottom baryons. The parameter $\alpha_s$ corresponds to the strong running coupling constant, which is written as
\begin{equation}
\alpha_{s}= \frac{\alpha_{s}(\mu_{0})}{1+\left(\frac{33-2n_{f}}{12 \pi}\right) \alpha_{s}(\mu_{0}) ln \left(\frac{m_{1}+ m_{2}+ m_{3}}{\mu_{0}}\right)}
\end{equation}
In above equation, the value of $\alpha_s$ at $\mu_0$ = 1 GeV is considered 0.6 as shown in Table \ref{tab:table4}.

The first order correction $V^{(1)}(x)$ can be written as
\begin{equation}
V^{(1)}(x)= - C_{F}C_{A} \frac{\alpha_{s}^{2}}{4 x^{2}}
\end{equation}
The parameters $C_F=2/3$ and $C_A=3$ are the Casimir charges of the fundamental and adjoint representation.\\

The spin dependent part $V_{SD}(x)$ is given as
\begin{eqnarray}
V_{SD}(x)= V_{SS}(x)(\vec{S_{\rho}}.\vec{S_\lambda})
+ V_{\gamma S}(x) (\vec{\gamma} \cdot \vec{S})&&  \nonumber \\ + V_{T} (x)
\left[ S^2-\frac{3(\vec{S }\cdot \vec{x})(\vec{S} \cdot \vec{x})}{x^{2}} \right]
\end{eqnarray}

The spin dependent potential, $V_{SD}(x)$ contains three types of the interaction terms \cite{12}, such as the spin-spin term $V_{SS} (x)$, the spin-orbit term $V_{\gamma S}(x)$ and tensor term $V_{T}(x)$ described as \cite{94,z}. Here $\vec{S}=\vec{S_{\rho}}+\vec{S_{\lambda}}$ where $\vec{S_{\rho}}$ and $\vec{S_{\lambda}}$ are the spin vector associated with the $\vec{\rho}$ and $\vec{\lambda}$ variables respectively. The coefficient of these spin-dependent terms of above equation can be written in terms of the vector, $V_V(x)$=$\frac{\tau}{x}$, and scalar, $V_S(x)$=$\beta x$, parts of the static potential as

\begin{equation}
V_{\gamma S} (x) = \frac{1}{2 m_{\rho} m_{\lambda}x}  \left(3\frac{dV_{V}}{dx} -\frac{dV_{S}}{dx} \right)
\end{equation}
\begin{equation}
V_{T}(x)=\frac{1}{6 m_{\rho} m_{\lambda}} \left(3\frac{d^{2}V_{V}}{dx^{2}} -\frac{1}{x}\frac{dV_{V}}{dx} \right)
\end{equation}

\begin{equation}
V_{SS}(x)= \frac{1}{3 m_{\rho} m_{\lambda}} \bigtriangledown^{2} V_{V}
\end{equation}

   The baryon masses are determined by the sum of the model quark masses plus kinetic
energy, potential energy and the spin dependent interaction as $M_B = \sum_i{m_i} + \langle H \rangle$.  We have numerically solved the six dimensional Schrodinger equation using Mathematica notebook \cite{Lucha1999}.

\section{Singly Bottom Baryon Spectra and Regge Trajectory}
The mass spectroscopy of single bottom baryons $\Lambda_b^0$, $\Sigma_b^{+,-}$, $\Xi_b^{-,0}$ and $\Omega_b^0$ have studied in the framework of Hypercentral Constituent Quark Model (hCQM). We have calculated the masses of these baryons for S, P, D and F states as presented in Table (\ref{tab:4} - \ref{tab:9}) where the tabular entries below A are masses without first order correction while below B are masses with first order correction. We have followed the n $^{(2S+1)} {L}_{J}$ usual notations for spectra of baryons except $L$ (angular momentum quantum number) is replaced by $\gamma$ (hyper-angular momentum quantum number) according to our model. We have considered all possible isospin splitting for the calculations of bottom baryons in all cases and the comparison of masses with other approaches are also tabulated.

$\Lambda_b^{0}$ was the first  experimentally known singly bottom baryon. The $\Lambda_{b}(5619)$ is the ground state, assigned to $J^P = \frac{1}{2}^{+}$. The first orbital excited states with $J^P = \frac{1}{2}^{-}$, $\frac{3}{2}^{-}$ are  $\Lambda_{b}(5912)$ and $\Lambda_{b}(5920)$ respectively. The computed mass spectra of $\Lambda_{b}^{0}$ baryon for 1S-6S, 1P-5P, 1D-4D and 1F-4F states are listed in Table~\ref{tab:4}. The 1P states, $\Lambda_{b} (\frac{1}{2}^{-})$ and $\Lambda_{b} (\frac{3}{2}^{-})$ are in accordance with experimental $\Lambda_{b}(5912)$ and $\Lambda_{b}(5920)$ states.

\par We computed the ground state as well as higher excited states with and without first order corrections for $\Sigma_{b}^+$ and $\Sigma_{b}^{-}$ baryons are listed in Table \ref{tab:7}. The PDG (2016) has listed  $\Sigma_{b}(5811)^+$ and $\Sigma_{b}(5816)^{-}$ state with $J^P = \frac{1}{2}^{+}$ and the states $\Sigma_{b}(5832)^{+*}$, $\Sigma_{b}(5835)^{-*}$ with $J^P = \frac{3}{2}^{+}$. Our results are in good agreement with the PDG (average) values as well as in accordance with the lattice result of
$m_{\Sigma_{b}}({1}/{2}^{+})$; 5856(56)(27) \cite{brown} \& 5795(16) \cite{120},
$m_{\Sigma_{b}}({3}/{2}^{+})$;5877(55)(25) \cite{brown} \& 5842(26) \cite{120}\\
\par PDG (2016) has listed $\Xi_{b}(5790)^{-}$, $\Xi_{b}(5790)^{0}$ and $\Xi_{b}(5945)^{0}$ as the lowest states. They are assigned as ground state with $J^{P}= \frac{1}{2}^{+}$ for first two and $J^{P}= \frac{3}{2}^{+}$ for third. Recent experiments show the values of $\Xi_{b}$ as(5.792 $\pm$ 0.0024) GeV \cite{CDF2007} and (5.774 $\pm$ 0.013) GeV \cite{103}. The present results as listed in Table \ref{tab:11}) are in good agreement with the experimental values as well as with the lattice calculations given by
$m_{\Xi_{b}}({1}/{2}^{+})$; 5771(41)(24) \cite{brown} \& 5781(17)(16) \cite{120}
$m_{\Xi_{b}}({3}/{2}^{+})$; 5960(47)(25) \cite{brown} \& 5950(21)(19) \cite{120}\\

\par $\Omega_{b}^-$ is  made up of two strange and one bottom quark ($\textit{ssb}$). The calculated mass spectra for radial (1S-5S) and orbital excited states (1P-5P,1D-2D and 1F) are given in Table \ref{tab:9}.  Our results are in good agreement with other model predictions as well as with the lattice result of
$m_{\Omega_{b}}({1}/{2}^{+})$; 6056(47)(20) \cite{brown} \& 6006(10)(20) \cite{120};
$m_{\Omega_{b}}({3}/{2}^{+})$; 6085(47)(20) \cite{brown} \& 6044(18)(20) \cite{120}\\


Our predicted 1P states of $\Lambda_{b}$ found to lie between 5.9 to 6.0 GeV and its 1D states within 6.20 to 6.23 GeV. The 1P states of $\Sigma_{b}$, $\Xi_b$ and $\Omega_b$ baryons found to lie 6.1 to 6.13 GeV, 6.12 to 6.14 GeV and 6.31 to 6.34 GeV respectively. The spectral 1D states of bottom baryons are predicted in the range of 6.34 to 6.40 GeV in the case of $\Sigma_b$ and $\Xi_b$ while that for $\Omega_b$ lie in the range of 6.54 to 6.6 GeV.\\

\begin{figure*}
\centering
\begin{minipage}[b]{0.47\linewidth}
\includegraphics[scale=0.26]{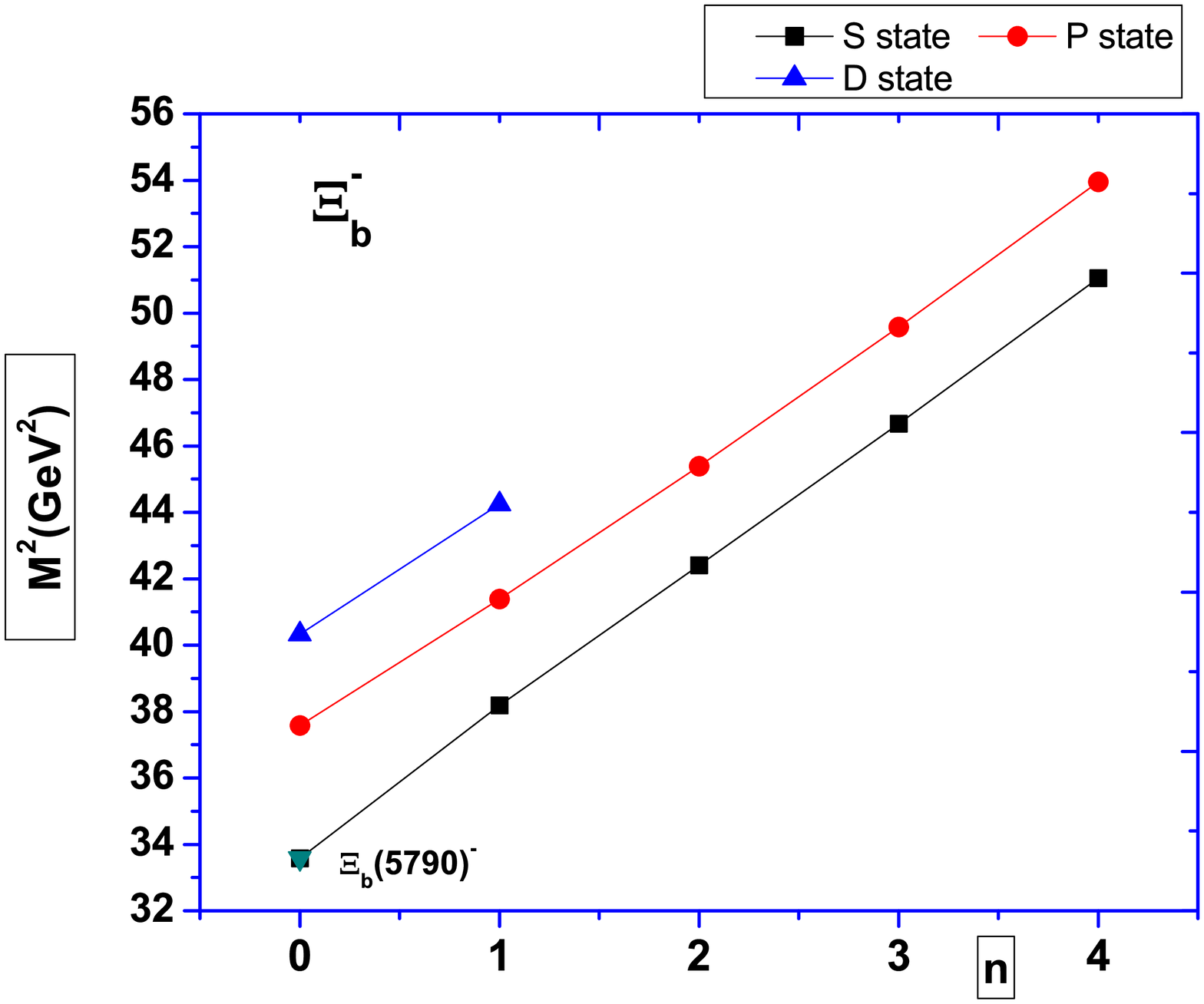}
\label{fig:minipage1}
\end{minipage}
\quad
\begin{minipage}[b]{0.47\linewidth}
\includegraphics[scale=0.26]{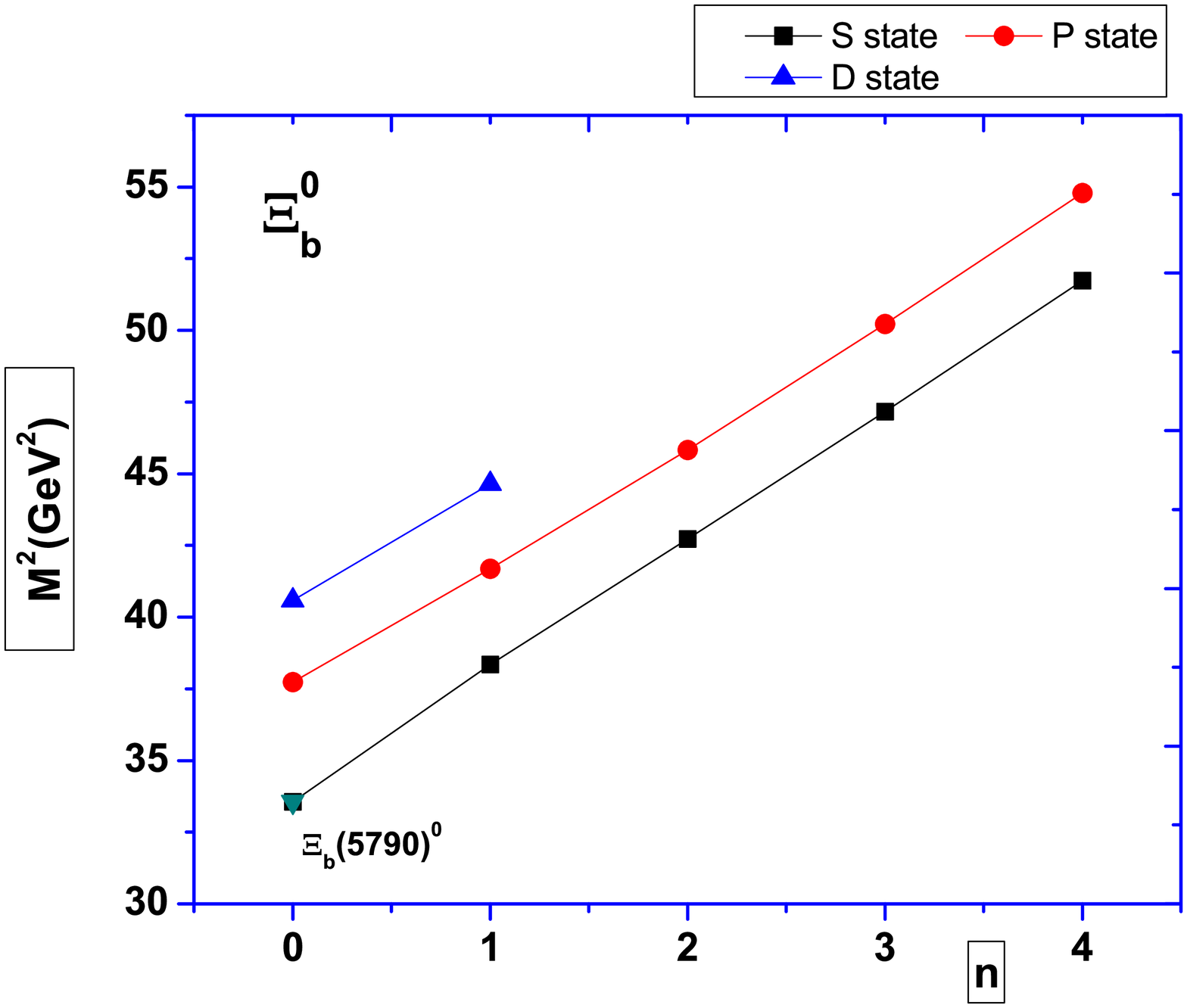}
\label{fig:minipage2}
\end{minipage}
\caption{\label{fig:3}Regge Trajectories ($M^{2}$ $\rightarrow$ n) for
$\Xi_{b}$ baryons.}
\end{figure*}

Regge theory provided an important relation between high energy scattering and spectrum of particles and resonances. It is a successful fundamental theory of strong interactions at very high energies and still an indispensable tool in phenomenological studies. One of the most distinctive features of Regge theory
are the Regge trajectories. Regge trajectories are directly related with mass spectrum of hadrons. Using the mass spectra computed for the singly bottom baryons, we construct the Regge trajectories in (n, $M^{2}$) planes [See Figs.~\ref{fig:1}-~\ref{fig:3}]. We use
 \begin{equation}
n=C M^2+ C_{0}
\end{equation}
\noindent where, $C$ and $C_{0}$ are slope and intercept respectively and n is the principal quantum number. In trajectories, the S, P and D state masses are corresponds to $J^{P}$= $\frac{1}{2}^{+}$,$\frac{1}{2}^{-}$ and $\frac{5}{2}^{+}$. Every singly bottom baryons have experimental known values for $J^P= \frac{1}{2}^{+}$. We include known experimental states $\Lambda_{b}^{0}(5619)$, $\Lambda_{b}^{0}(5912)$, $\Sigma_{b}^{+}(5811)$, $\Sigma_{b}^{-}(5816)$, $\Xi_{b}^{-}(5790)$, $\Xi_{b}^{-}(5790)$ and $\Omega_{b}^{-}(6048)$  and these experimental points fit well to the corresponding Regge trajectories. The calculated masses of present states fit very well to the linear trajectories and they are almost parallel and equidistant. Many of the  excited  states are still unknown experimentally for the case of singly bottom baryons. The Regge trajectories can provide guidelines to identify baryon resonances that would be seen in future experiments.

\begin{figure*}
\centering
\begin{minipage}[b]{0.47\linewidth}
\includegraphics[scale=0.26]{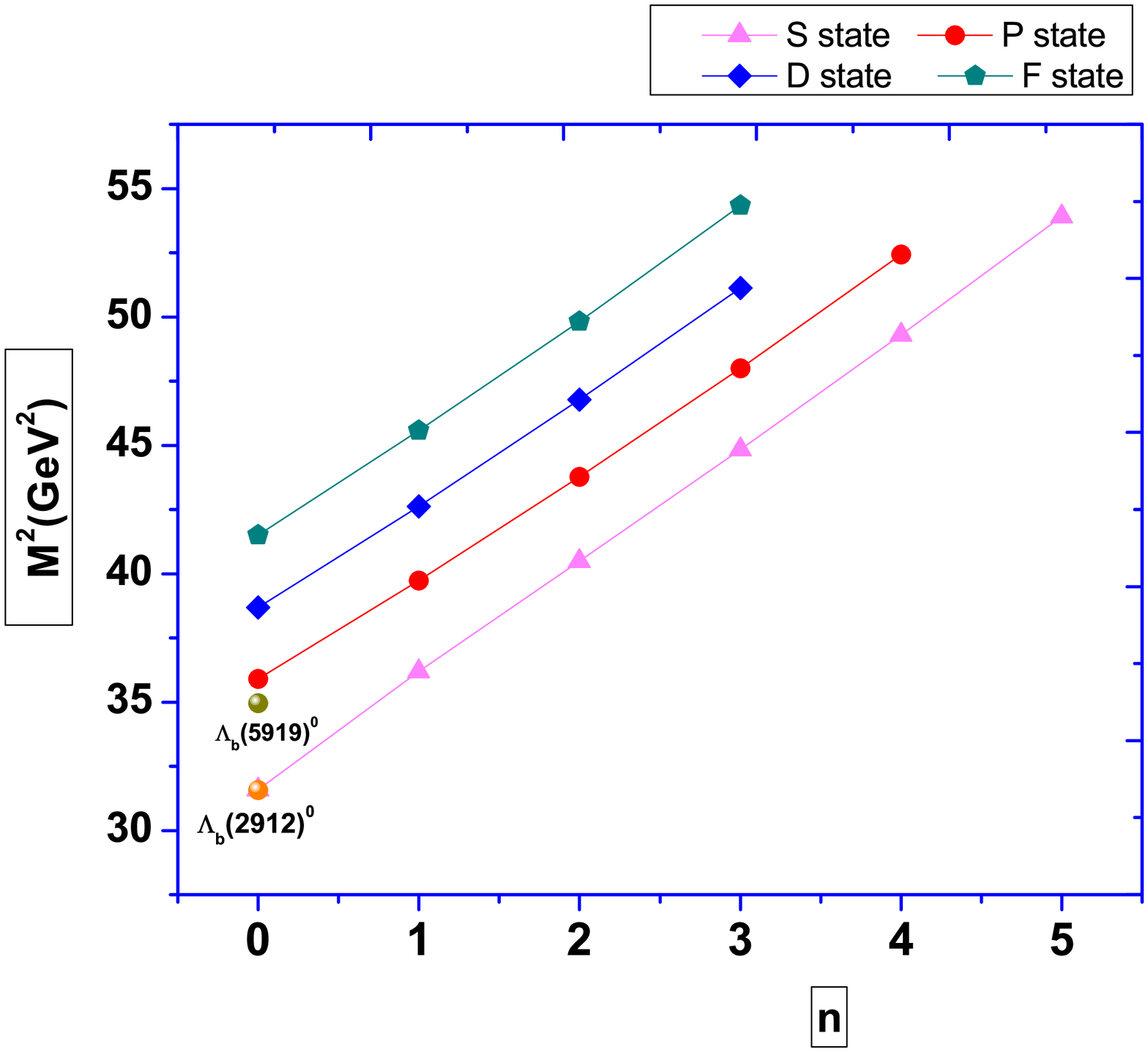}
\label{fig:minipage1}
\end{minipage}
\quad
\begin{minipage}[b]{0.47\linewidth}
\includegraphics[scale=0.26]{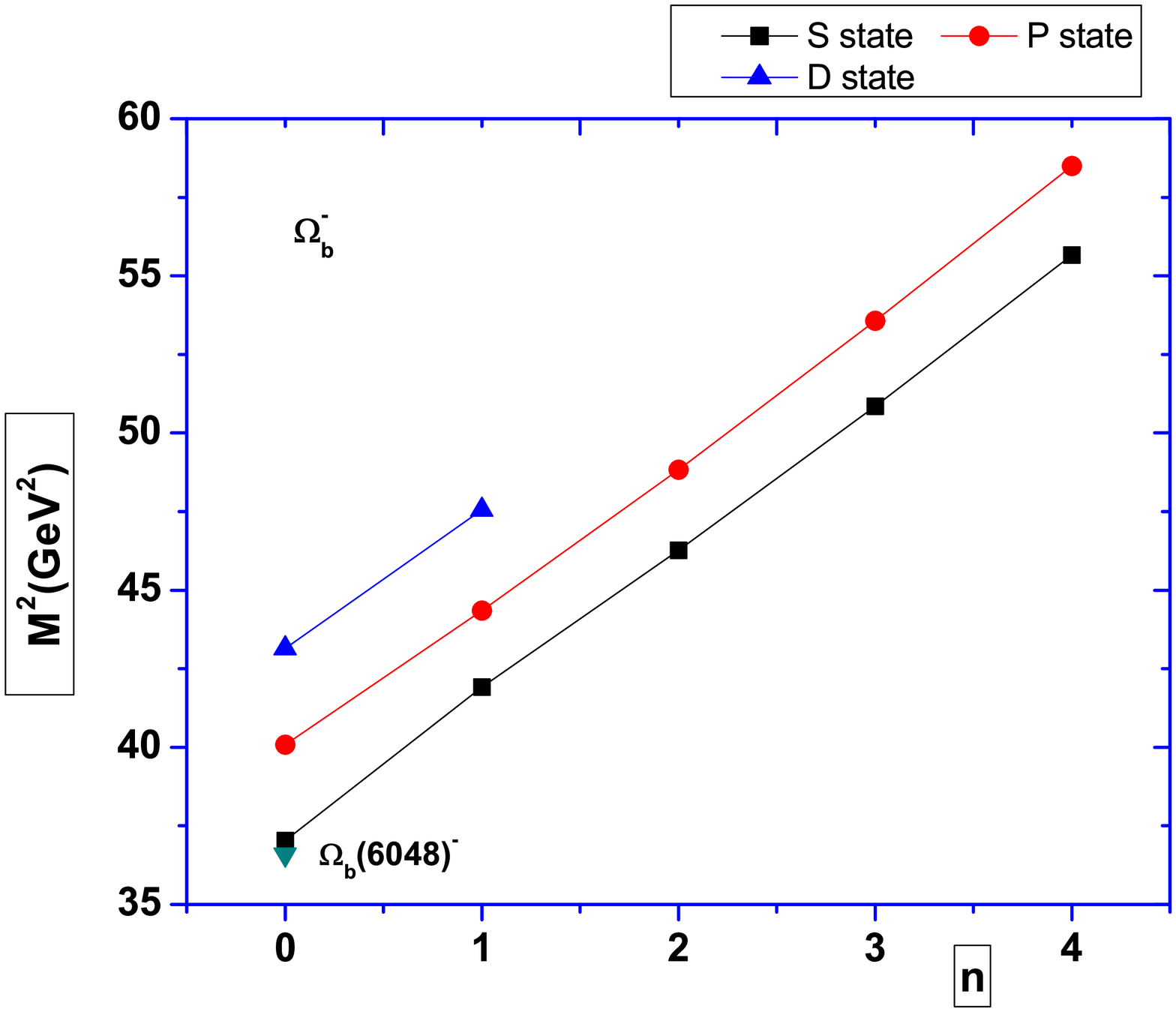}
\label{fig:1}
\end{minipage}
\caption{\label{fig:1} Regge Trajectory ($M^{2}$ $\rightarrow$ n) for
$\Lambda_{b}^{0}$(left) and $\Omega_{b}^{-}$ (right) baryons. Available Experimental data are given with particle name.}
\end{figure*}

\begin{figure*}
\centering
\begin{minipage}[b]{0.47\linewidth}
\includegraphics[scale=0.26]{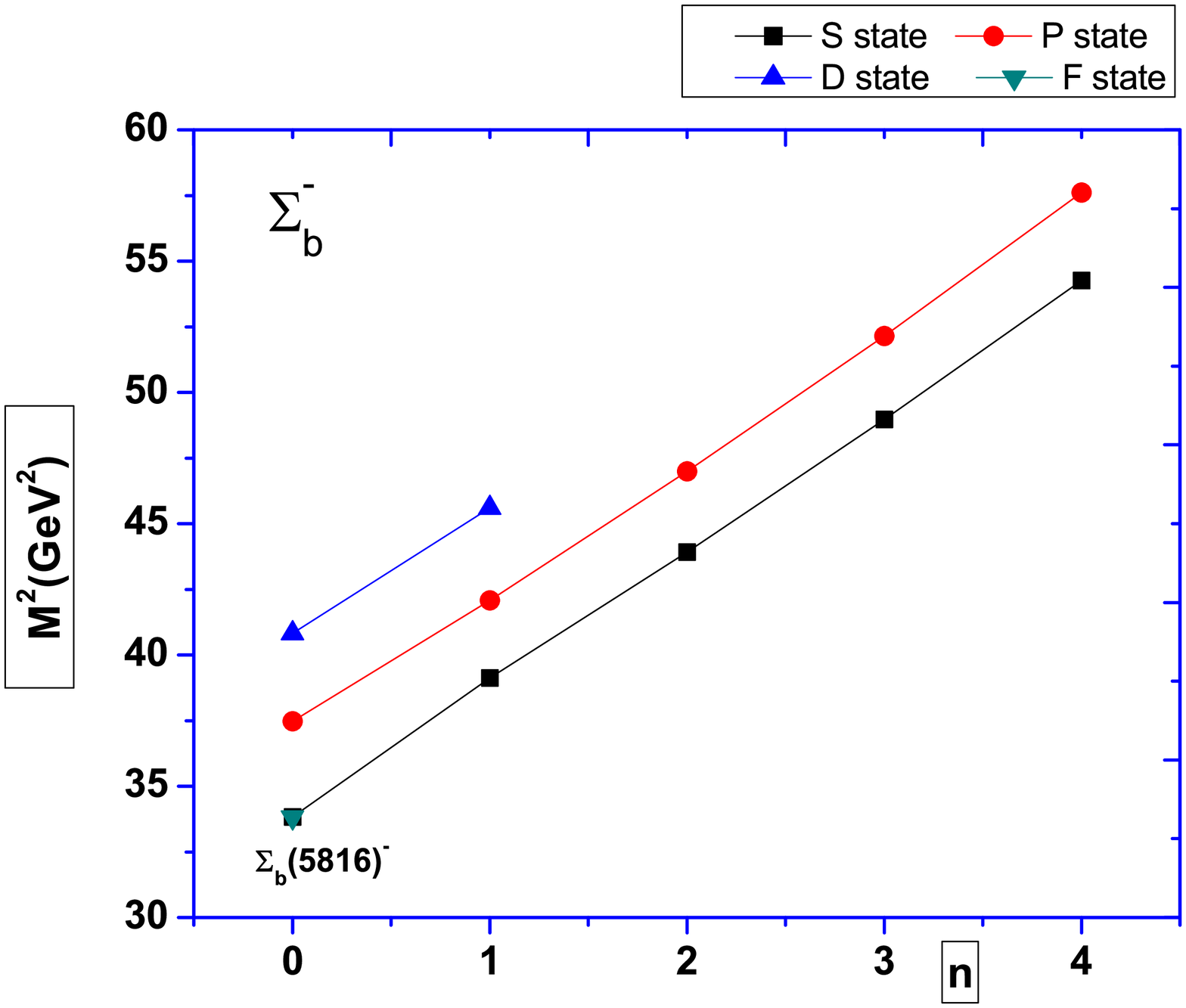}
\label{fig:minipage1}
\end{minipage}
\quad
\begin{minipage}[b]{0.47\linewidth}
\includegraphics[scale=0.26]{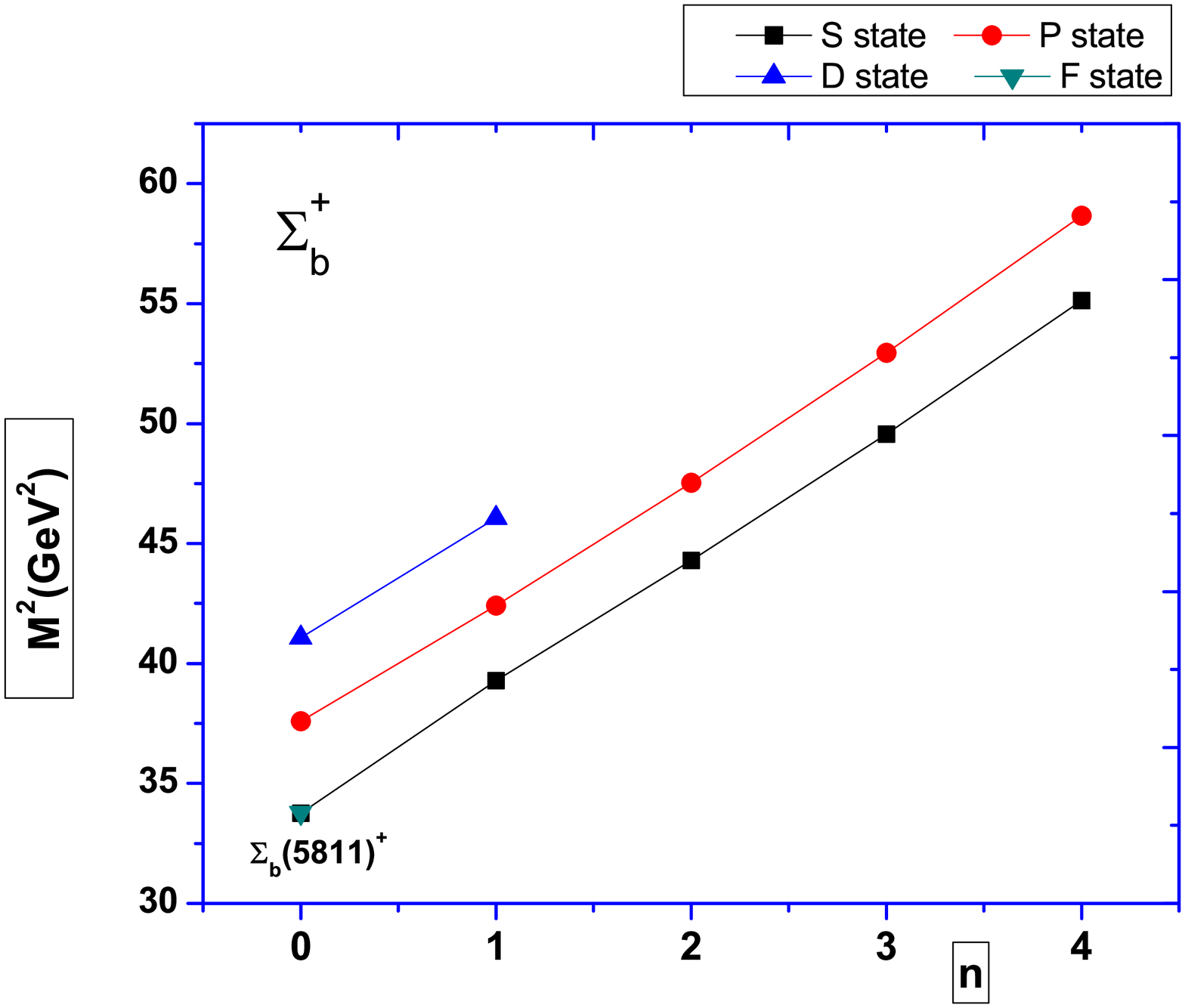}
\label{fig:minipage2}
\end{minipage}
\caption{\label{fig:2} Regge Trajectories ($M^{2}$ $\rightarrow$ n) for
$\Sigma_{b}$ baryons.}
\end{figure*}

\section{Semi-electronic decays of $\Xi_{b}$ and $\Omega_{b}$ baryons }

In weak decays, the heavy quark(c or b) acts as spectator and the strange quark inside heavy hadron decays in weak interaction \cite{sven, Hai2, Hai}. These could be possible in semi-electronic, semi-muonic and non leptonic decays of the heavy baryons and mesons. In this section, we discuss, Semi-electronic decays of heavy bottom baryons $\Omega_{b}$ and $\Xi_{b}$ using our spectral parameters. The strange quark in $\Xi_{b}$ and $\Omega_{b}$ undergo weak transitions and the differential decay rates for exclusive semi-electronic decays are given by \cite{sven},
\begin{eqnarray}\label{eq:11}
\frac{d \Gamma}{dw} = \frac{G_{F}^{2} M^{5} \vert V_{CKM}\vert^{2}}{192 \pi^{3}} \sqrt{w^{2}-1} P(w)
\end{eqnarray}
where $P(w)$ contains the hadronic and leptonic tensor. After evaluating the integration over $w$=1 in the hadronic form factors, we calculate semi-electronic decay as given below. \\
For the final state with ``$\Lambda_b$" baryon,
\begin{equation}\label{eq:12}
\Gamma^{\frac{1}{2}^{+} \rightarrow \frac{1}{2}^{+}}_{0^{+} \rightarrow 0^{+}} = \frac{G_{F}^{2}\vert V_{CKM}\vert^{2}}{60 \pi^{3}} (M-m)^{5}
\end{equation}
For the final state with ``$\Xi_b$" baryon,
\begin{equation}\label{eq:13}
\Gamma^{\frac{1}{2}^{+} \rightarrow \frac{1}{2}^{+}}_{1^{+} \rightarrow 1^{+}} = \frac{G_{F}^{2}\vert V_{CKM}\vert^{2}}{15 \pi^{3}} (M-m)^{5}
\end{equation}

\begin{table}
\caption{\label{tab:12}Semi-electronic decays in $s \rightarrow u$ transition for bottom baryons are listed.}
 \resizebox{\textwidth}{!}{
\begin{tabular}{ccccccccc}
\hline\hline
Mode & $J^{P} \rightarrow J'^{P{'}}$ & $s^{l} \rightarrow s'^{l'}$& {$\bigtriangleup m$ (GeV)}& {Decay Rates
 (GeV)} & \cite{sven}\\
\hline
$\Xi_b^- \rightarrow \Lambda_b^{0} e^{-} \bar{\nu}$ & $\frac{1}{2}^{+} \rightarrow \frac{1}{2}^{+}$& $0 \rightarrow 0$ &0.174&  $5.928\times 10^{-19}$& $6.16\times 10^{-19} $\\
$\Omega_b^- \rightarrow \Xi_b^{0} e^{-} \bar{\nu}$ & $\frac{1}{2}^{+} \rightarrow \frac{1}{2}^{+}$ & $1 \rightarrow 0$ &0.255&  $4.007\times 10^{-18} $ & $4.05\times 10^{-18} $   \\
$\Omega_b^- \rightarrow \Xi_b^{*0} e^{-} \bar{\nu}$ & $\frac{1}{2}^{+} \rightarrow \frac{3}{2}^{+}$ & $1 \rightarrow 1$ &0.101& $1.675\times 10^{-26} $ &$3.27\times 10^{-28} $\\
\hline\hline
\end{tabular}}
\end{table}

Where $G_{F}$ is the Fermi Coupling constant and the value of $G_{F}$ = 1.16$\times10^{-5}$     $GeV^{-2}$, $V_{CKM}$ is the Cabibbo-Kobayashi-Maskawa matrix and we have taken the value of $V_{CKM}$ = 0.225, $\Delta m$ = $M-m$, is the mass difference between the initial and final
state of baryons.  The superscript of $\Gamma$ in $Eqn$.(\ref{eq:11}), (\ref{eq:12}) and (\ref{eq:13}) indicates spin parity transition ($J^{P} \rightarrow J'^{P{'}}$) of baryon , while the subscripts of $\Gamma$ indicate spin parity transition ($s^l \rightarrow s'^{{l}{'}}$) of light degrees of freedom. Semi-electronic decays in $s \rightarrow u$ transition for bottom baryons ($\Xi_b^{0}$ and $\Omega_b^{0}$) are tabulated in Table~\ref{tab:12}. The initial and final total angular momentum (J) and parity (P), total spin $s_{l}$ of the light degree of freedom and the mass difference $\bigtriangleup m$ = M - m for Baryons are listed in second, third and forth column of the table. We have also compared our results with Ref. \cite{sven}.
\begin{figure*}
\centering
\begin{minipage}[b]{0.47\linewidth}
\includegraphics[scale=0.29]{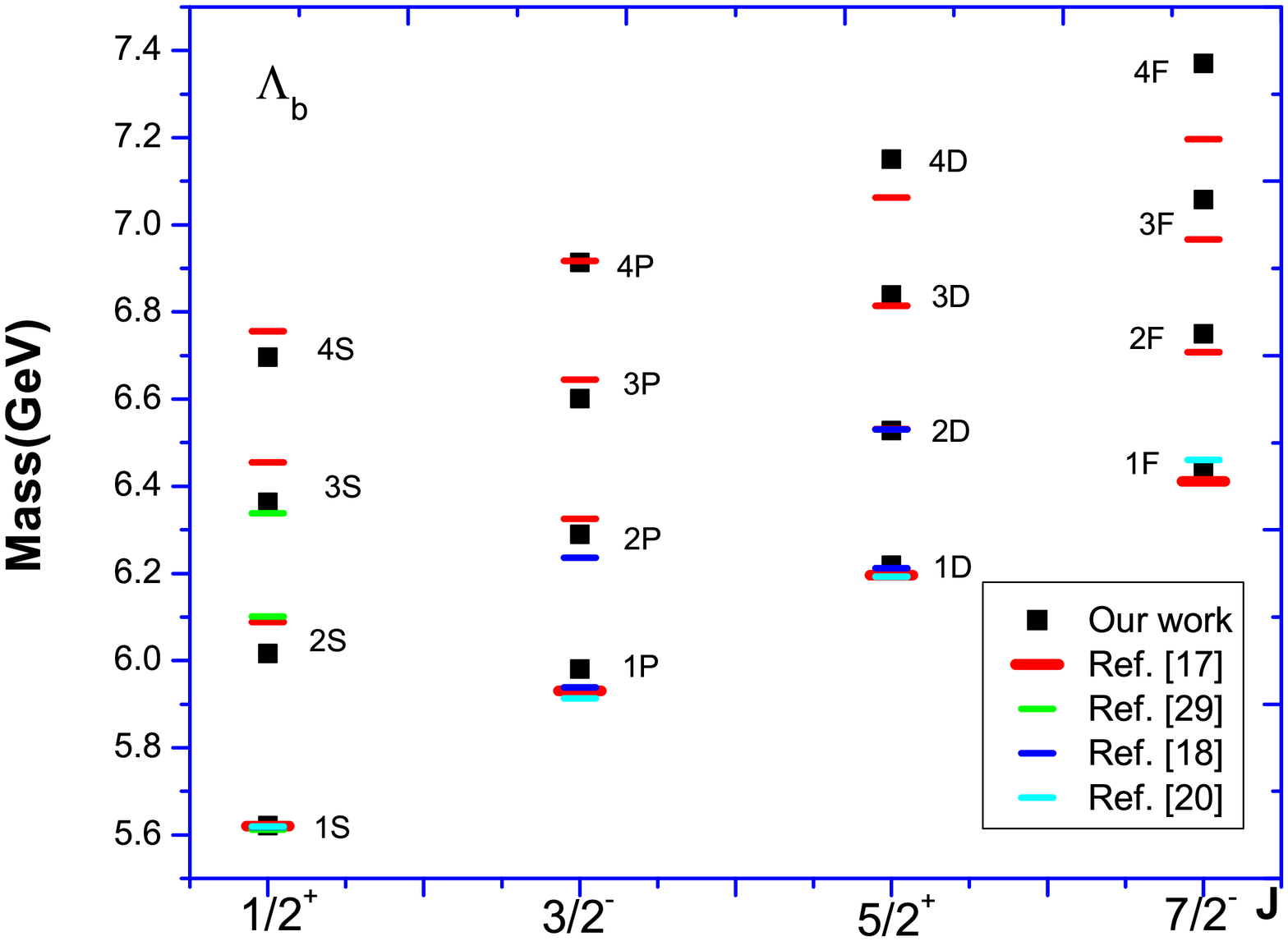}
\label{fig:minipage1}
\end{minipage}
\quad
\begin{minipage}[b]{0.47\linewidth}
\includegraphics[scale=0.29]{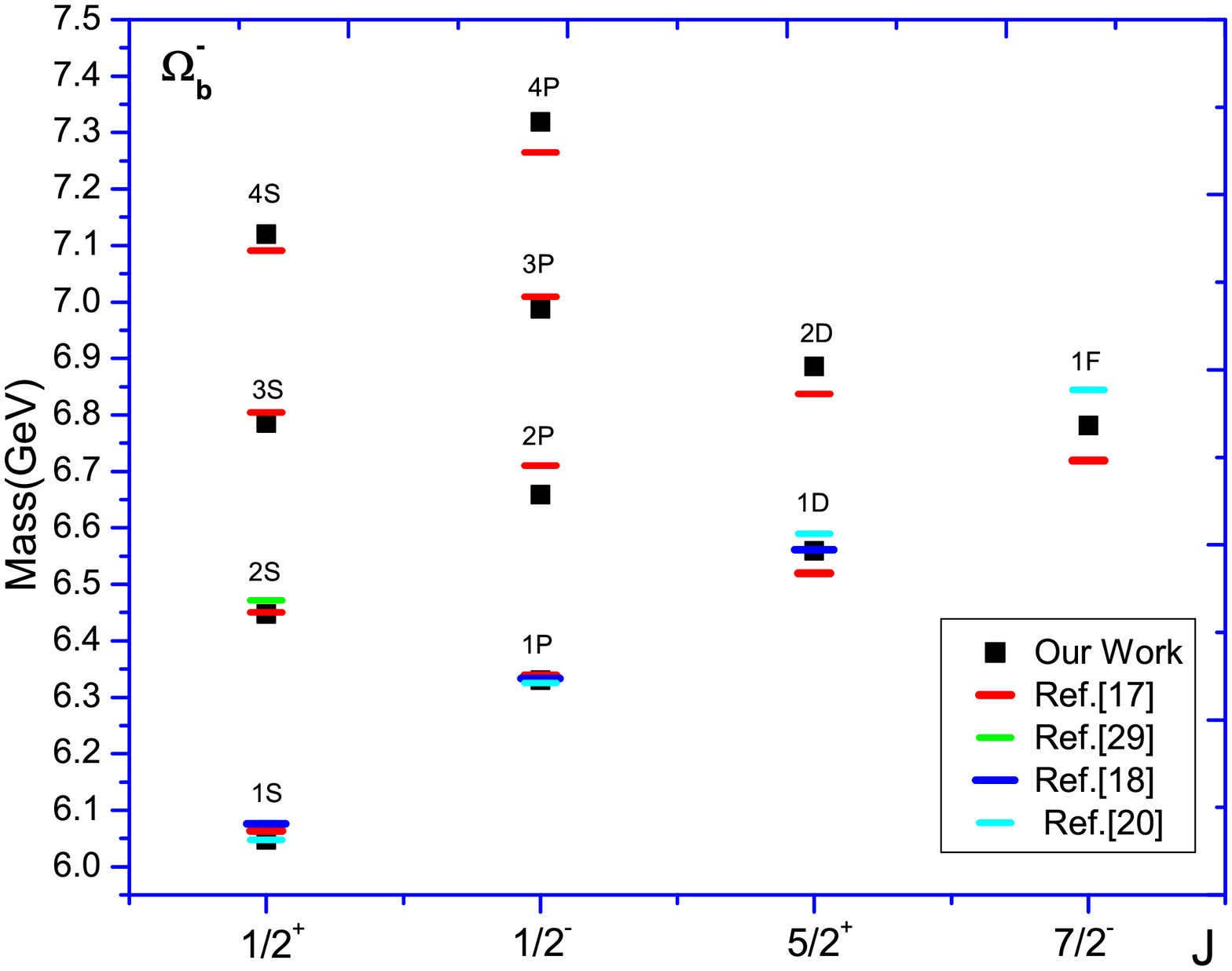}
\label{fig:minipage2}
\end{minipage}
\caption{\label{fig:4}Comparisons of excited bottom baryons masses with other prediction for $\Lambda_b$ and $\Omega_b$.}
\end{figure*}

\begin{figure*}
\centering
\begin{minipage}[b]{0.47\linewidth}
\includegraphics[scale=0.29]{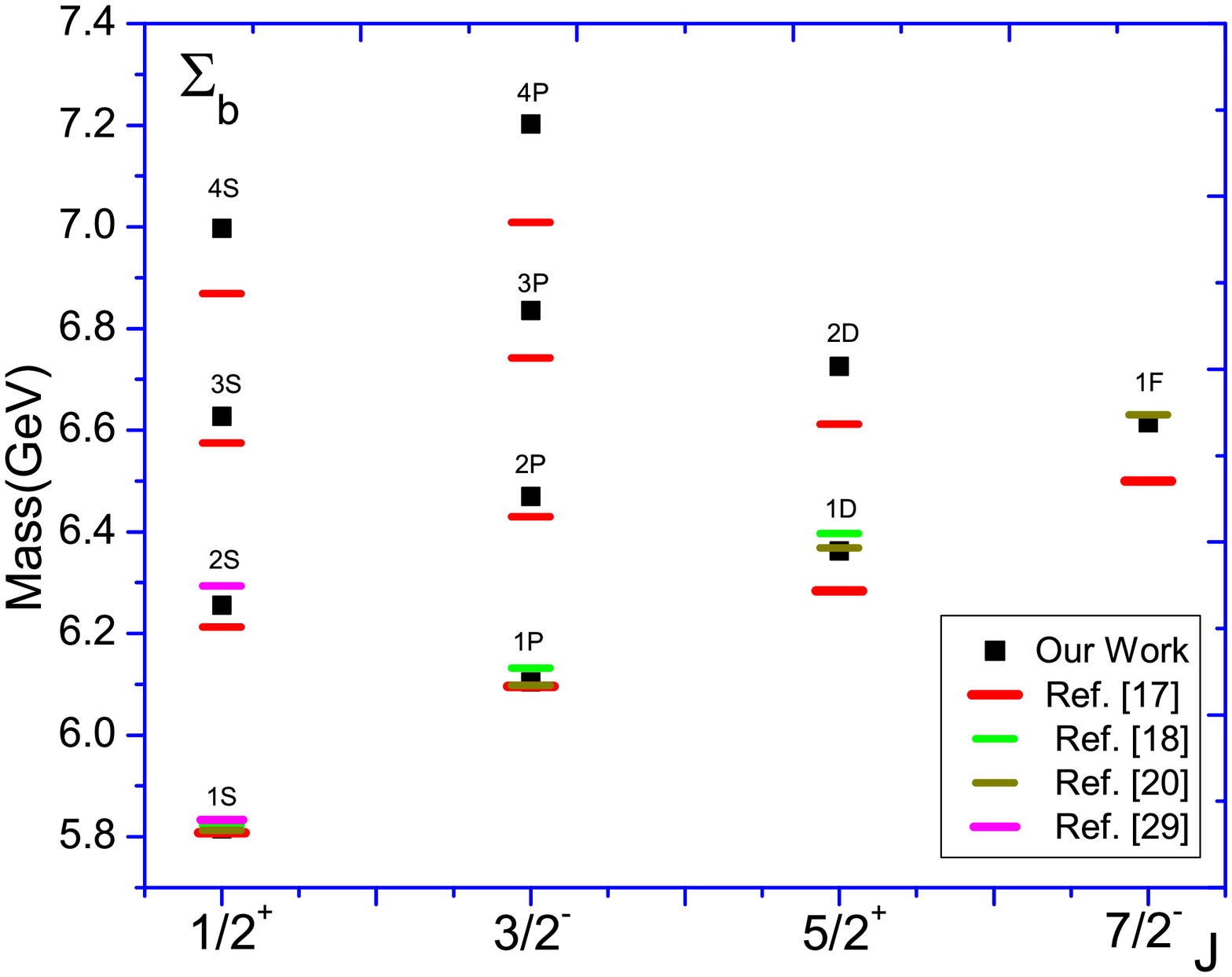}
\label{fig:minipage1}
\end{minipage}
\quad
\begin{minipage}[b]{0.47\linewidth}
\includegraphics[scale=0.29]{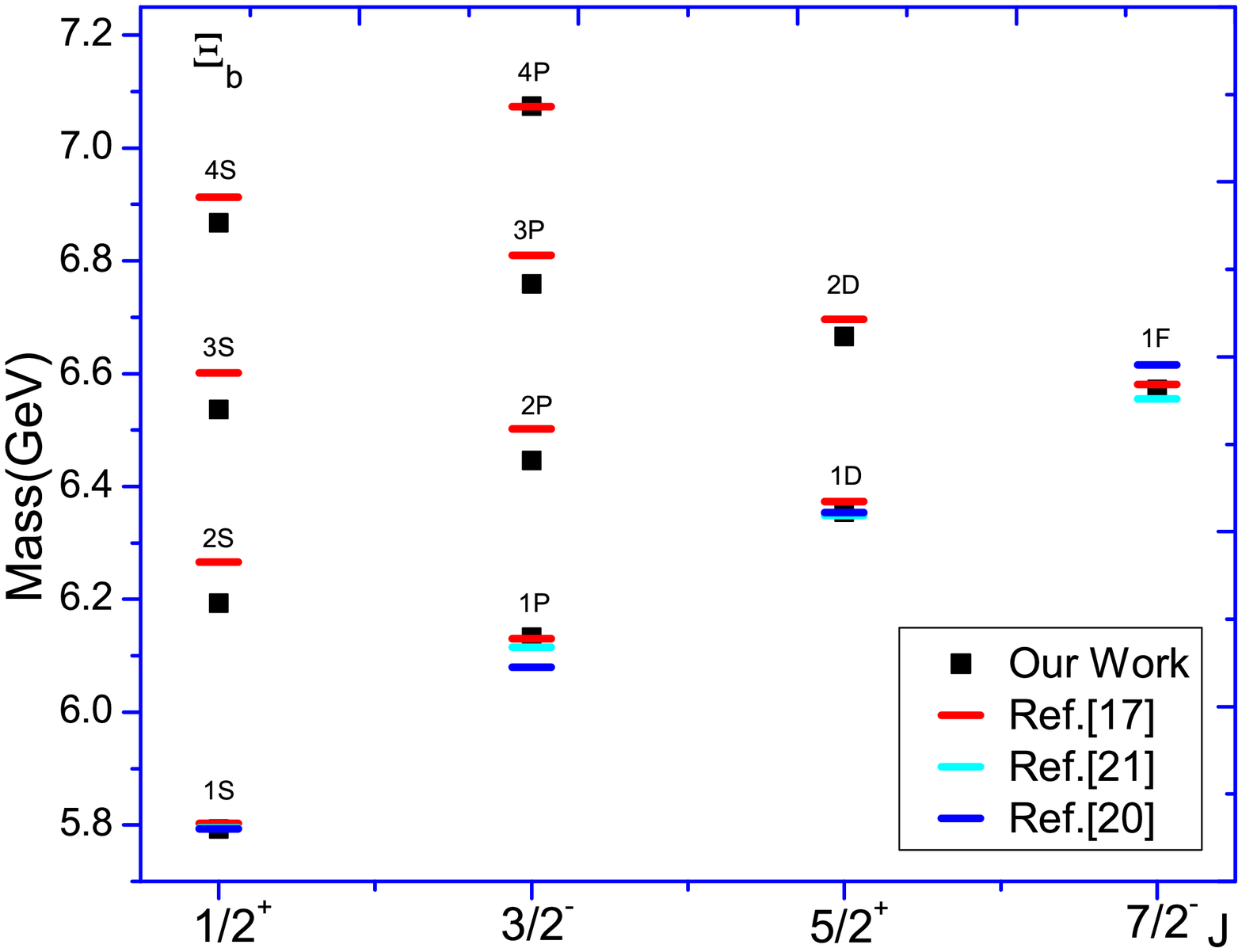}
\label{fig:minipage2}
\end{minipage}
\caption{\label{fig:5}Comparisons of excited bottom baryons masses with other prediction for $\Sigma_b$ and $\Xi_b$.}
\end{figure*}

\section{Conclusion}
 We have calculated the mass spectra of $\Lambda_b^0$, $\Sigma_b^+$, $\Sigma_b^0$, $\Xi_b^0$, $\Xi_b^-$ and $\Omega_b^0$ baryons using Hypercentral Constituent quark model with an effective three body interactions of the type hyper Coulomb plus linear potential.  The whole mass spectra of these singly bottom baryons are listed for few radial and orbital excited states. For the sake of simplicity and better understanding of the results, the mass spectra are shown in energy level diagram of the baryons with a particular $J^{P}$ value in Figs.~\ref{fig:4}-~\ref{fig:5}. The ground states of $\Lambda_b$, $\Sigma_b$, $\Xi_b$ and $\Omega_b$ are well-known by experimental and Lattice-QCD results. Thus, 1S($\frac{1}{2}^{+}$) state values are overlapping with each other in all cases[See Figs.~\ref{fig:4}-~\ref{fig:5}]. By observing these figures, we can conclude these points.
 \begin{enumerate}
\item{In the study of $\Lambda_b$ baryon, the states which are reasonably close with other predictions are 2S-3S(with \cite{Yamaguchi2015}), 2P (with \cite{yoshida,ebert2011}), 1D (with \cite{ebert2011,yoshida,kwei}), 2D(with \cite{ebert2011,yoshida}) and 1F(with \cite{ebert2011,kwei}). Therefore, the closest states with our predictions(A) are:
Ref. \cite{ebert2011} show 8, 4, 22 and 4 MeV difference in 4P, 1D, 2D, 1F states, Ref. \cite{yoshida} show 17, 22 MeV difference in 1P, 2D states.}

\item{In $\Omega_b$ baryon, 2S-4S, 1P-3P and 1D states  are in accordance with \cite{ebert2011}; 1P and 1D states are in accordance with \cite{yoshida,kwei}. Our prediction of the 1F state is in between the prediction of Ref. \cite{ebert2011, kwei} so that the experimental outcomes will decide the actual range. Therefore, the closest states with our predictions(A) are:
Ref \cite{ebert2011} show 2, 18, 29, 19, 25 difference in 2S, 3S, 4S, 1P, 3P states;  \cite{yoshida} show 15, 7 difference in 1P, 1D states. }
\item{In $\Sigma_b$, comparisons of states are as follows; 2S (with \cite{ebert2011, Yamaguchi2015}, 1P(with \cite{ebert2011,kwei,yoshida}), 1D(\cite{yoshida, kwei}) and 1F(with \cite{kwei}). Therefore, the closest states with our predictions(A) are: \cite{ebert2011} show 9 MeV difference in 1P state; \cite{yoshida} show 7, 34 MeV difference in 1P, 1D state; \cite{kwei} 6, 15 MeV difference with 1D, 1F states.}
\item{In the study of $\Xi_b$, Refs. \cite{ebert2011,kwei, 118} have performed the 1P, 1D and 1F states. These states are very much close to our predictions while radial excited states of $\Xi_b$ are fairly match with our predictions.
}

 \end{enumerate}
This detailed evaluation of mass spectra of singly heavy baryon will definitely help other approaches and experiments to find resonances in theoretically predicted range. The regge trajectories are also very useful to obtain unknown quantum number and respective $J^P$ values of bottom baryons. Mass study will definitely help to understand the nearly available resonances belong to singly bottom baryons.  The semi-electronic decays are also calculated for $\Xi_b$ and $\Omega_b$ baryons. Where, the $\Omega_b^- \rightarrow \Xi_b^{*0} e^{-} \bar{\nu}$ decay rate is disagree with \cite{sven}.
\par After successful implementation of this scheme to the singly heavy baryons (both charm and bottom) as well as doubly heavy baryons, we would like to calculate the decay rates of heavy baryons in near future. The decay properties are very important to understand the dynamics of baryons so our next attempt would be the study of various decay properties.\\

\textbf{Acknowledgments:}
One of the author A. K. Rai acknowledges the financial support extended by DST, India  under SERB fast track scheme SR/FTP /PS-152/2012.\\




\end{document}